# Analytical formula of Free Electron Laser exponential gain for non-resonant electron beam

Qika Jia
National Synchrotron Radiation Laboratory,
University of Science and Technology of China,
Hefei, Anhui 230029, China

***Abstract*—**The FEL gain formulas for non-resonant case are studied. For the mono-energetic and non-resonant electron beam, the exact expression of the solution of the FEL characteristic cubic equation is obtained with a form much more simple than that in the literatures，and the gain length as the function of the detuning parameter is explicitly given, then the gain for different detuning parameter and from low to high can be easily calculated. A simplified approximation formula is also given for the exponential gain calculation in the non-resonant case. For the case of the electron beam with an energy spread, the solution of the characteristic cubic equation is given explicitly for rectangular energy distribution and Lorentz distribution, respectively. Moreover the explicit expression also can be used for the solution of the characteristic cubic equation including the impact of the space charge. The transition from the low gain to the high gain is analyzed. The variations of the gain bandwidth and of the detuning parameter for the maximum gain are demonstrated. The applicable ranges of the small signal gain formula and the exponential gain formula are analyzed.

## I, INTRODUCTION

The optical field gain is one of most important parameters in free electron laser (FEL). In the linear regime, the FEL optical field is under saturation, usually it has a low gain per pass in oscillator mode and a high gain in single-amplifying mode. The level of the optical field gain of FEL is determined by the intensity of the electron current density and the length of the undulator, the effects of the two factors can be characterized by the number of the gain length in the length of the undulator. The gain length is defined as the e-folding length that the optical power increased by a factor of e=2.718….

Though analytical gain formulas are convenient to calculation and analysis comparing with the numerical method, but they exist only for special cases of FEL gain calculation. Typically there are the small signal gain formula in the low gain regime and the exponential gain formula in the high gain regime, the former is the initial gain in the multi-amplifying of oscillator FEL and the later is the gain in single-amplifying high gain FEL. The existing exponential gain formula is only for the electron beam at the resonant energy. In this paper we consider the exponential gain for the non-resonant case, including the mono-energetic electron beam but non-resonant and the electron beam with an energy spread. We try to give some analytical expressions of the gain, and we also analyze the transition from the range of the small signal gain to the range of the exponential gain.

The paper is organized as following: in the second section we give a brief review of the existing analytic theory of the exponential gain; then in the third section the general non-resonant case will be analyzed; the case of the electron beam with an initial energy spread will be considered in the fourth section; and at last we give an analysis on the relation between the different gain formulas.

## II, THE EXISTING ANALYTIC THEORY OF THE EXPONENTIAL GAIN[1-5]

The optical field in the linear regime for the mono-energetic electron beam is given as

$$\tilde{a}_s(\hat{z}) = i\sum_{\substack{m=1\\ m\neq l,k}}^{3} \frac{\tilde{a}_{s0}e^{\hat{\mu}_m\hat{z}}}{\hat{\mu}_m(\hat{\mu}_m-\hat{\mu}_l)(\hat{\mu}_m-\hat{\mu}_k)} = \tilde{a}_{s0}\sum_{m=1}^{3}\frac{\hat{\mu}_m+i\hat{\eta}}{3\hat{\mu}_m+i\hat{\eta}}e^{\hat{\mu}_m\hat{z}} \tag{1}$$

where $\tilde{a}_{s0}$ is the initial dimensionless vector potential of the optical field, $\hat{z}=2k_u\rho z$ is scaled position along the undulator, $k_u$ is wave number of the undulator, ρ is FEL parameter, $\hat{\eta}=\delta\gamma/\gamma\rho$ is the scaled detuning parameter, $\delta\gamma/\gamma$ is the beam energy deviation from the resonance, $\hat{\mu}_m$ ($m$= 1,2,3) is the roots of the well-known characteristic cubic equation:

$$\hat{\mu}(\hat{\mu}+i\hat{\eta})^2 = i \tag{2}$$

When $\hat{\eta}<3/\sqrt[3]{4}$ =1.89, the cubic equation has one imaginary root corresponding to the oscillatory mode of the optical field, and two complex roots with the real parts of equal magnitude but opposite sign, corresponding to the exponential amplifying mode and the exponential decaying mode, respectively. (and if $\hat{\eta}$> 1.89, the cubic equation will have three imaginary roots). For a sufficiently long undulator the leading role is the exponential growth term corresponding to the root with positive real part.

The maximum gain rate occurs at the resonant energy. For this case, the detuning parameter $\hat{\eta}$=0, the three roots are $\hat{\mu}_{1,2,3}=(i\pm\sqrt{3})/2,\ -i$, then the gain formula can be given analytically

$$\frac{P}{P_0} = \frac{1}{9}\left\{1+4ch(\frac{\sqrt{3}}{2}\hat{z})[ch(\frac{\sqrt{3}}{2}\hat{z})+\cos(\frac{3}{2}\hat{z})]\right\} \approx \frac{1}{9}e^{\frac{z}{L_g}} \tag{3}$$

In above equation, the last approximation of exponential gain is valid for large rate of $L/L_g$ (see section V), $L$ is the length of undulator, $L_g=1/2k_u\sqrt{3}\rho$ is the power gain length.

For the case of non-resonance, if the detuning is small $|\hat{\eta}|$<<1, namely near the resonance, the standard approach (e.g. in Ref.[1]-[3]) is expanding the cubic equation near $\hat{\eta}$=0 to the second order in $\hat{\eta}$ as

$$\hat{\mu} = \hat{\mu}_0 - i\frac{2}{3}\hat{\eta} - \frac{1}{9\hat{\mu}_0}\hat{\eta}^2 \tag{4}$$

then the root corresponding to the exponential growth term is

$$\hat{\mu}_1 = \frac{\sqrt{3}}{2}[1-\frac{1}{9}\hat{\eta}^2] + \frac{i}{2}[1-\frac{4}{3}\hat{\eta}+\frac{1}{9}\hat{\eta}^2] \tag{5}$$

and the exponential gain of the optical field is

$$\tilde{a}_s/\tilde{a}_{s0} \propto e^{\sqrt{3}\hat{z}}e^{i[\frac{1}{2}-\frac{2}{3}\hat{\eta}]\hat{z}}e^{-\frac{\hat{\eta}^2}{4\sigma_\eta^2}[1-\frac{i}{\sqrt{3}}]} \tag{6}$$

where $\sigma_\eta^2 = 3\sqrt{3}/2\hat{z}$ is the gain bandwidth, follow from which the radiation bandwidth can be given as $\sigma_\omega/\omega = 2\rho\sigma_\eta$. It decreases with the increase of $z$ and goes to ρ when the optical field tends to saturation.

It should be pointed that actually besides the exponential factor in Eq. (6), the pre-exponential factor is also relevant to the detuning parameter.

**III, THE GENERAL NON-RESONANT CASE**

For the mono-energetic and non-resonant electron beam, the exact solution of the cubic equation (Eq.2) was given, but in a very complicated form (for example as in reference [1],[6]). By careful simplifying we get the exact solution with much more simple form as following:

$$\hat{\mu}_{1,2} = \pm(p-q) + i(\frac{p+q}{\sqrt{3}} - \frac{2\hat{\eta}}{3})$$

$$\hat{\mu}_3 = -i2(\frac{p+q}{\sqrt{3}} + \frac{\hat{\eta}}{3}) \tag{7}$$

$$p,q = \frac{\sqrt{3}}{2\sqrt[3]{4}}\left\{1 \pm \sqrt{1-4(\frac{\hat{\eta}}{3})^3}\right\}^{\frac{2}{3}}, \ (\hat{\eta} = \frac{\delta\gamma}{\gamma_r \rho} < \frac{3}{\sqrt[3]{4}})$$

From the root $\hat{\mu}_1$ we obtan the gain length explicitly as：

$$\frac{L_g}{L_{g0}} = \sqrt[3]{4} / \left\{[1+\sqrt{1-4(\frac{\hat{\eta}}{3})^3}\,]^{\frac{2}{3}} - [1-\sqrt{1-4(\frac{\hat{\eta}}{3})^3}\,]^{\frac{2}{3}}\right\} \tag{8}$$

where $L_{g0}$ is the gain length for the resonant case. From Eq.(8) it is obvious to the requirement of $\hat{\eta} < 3/\sqrt[3]{4}$ =1.89. Variation of the gain length with the detuning parameter (i.e.the electron energy) is shown as Figure 1, the solid curve is exactly same as in the literatures (e.g. in Ref [1],[2],[7]) where it was given by numerically, but here it's given by the formula. Therefore the gain length of general case can be calculated conveniently. At the resonance the detuning parameter equal to zero, and the gain length is the shortest, the growth rate reaches its maximum. The gain length increase sharply for the detuning parameter larger than zero and slowly for the detuning parameter smaller than zero. The result given by Eq. (5) is also plotted in Figure 1， it can be seen that it is only applicable for the case of $|\hat{\eta}|$<1.

By using Eq.(7) and Eq. (1), the variation of the gain with the detuning parameter can be calculated. The result for the case of near the saturation ($L$～$20L_g$) is shown in Figure 2. It is seen that for the high gain, the gain bandwidth $\Delta\hat{\eta}_{FWHM} \sim 1$, this agrees with the result from Eq. (6): $\sigma_{\hat{\eta}}$~1/2. From Figure 2 one also can see that the maximum gain is not at the resonance $\hat{\eta}$=0 as is the maximum growth rate (Figure 1), but is at $\hat{\eta}$ ≈0.12, and for the detuning parameter smaller than zero, the gain does not vary slowly as the gain length does (Fig.1). These are because that for the non-resonant case the non-exponential factor in the optical field expression (Eq. 1) is also related to the detuning as we pointed previously.

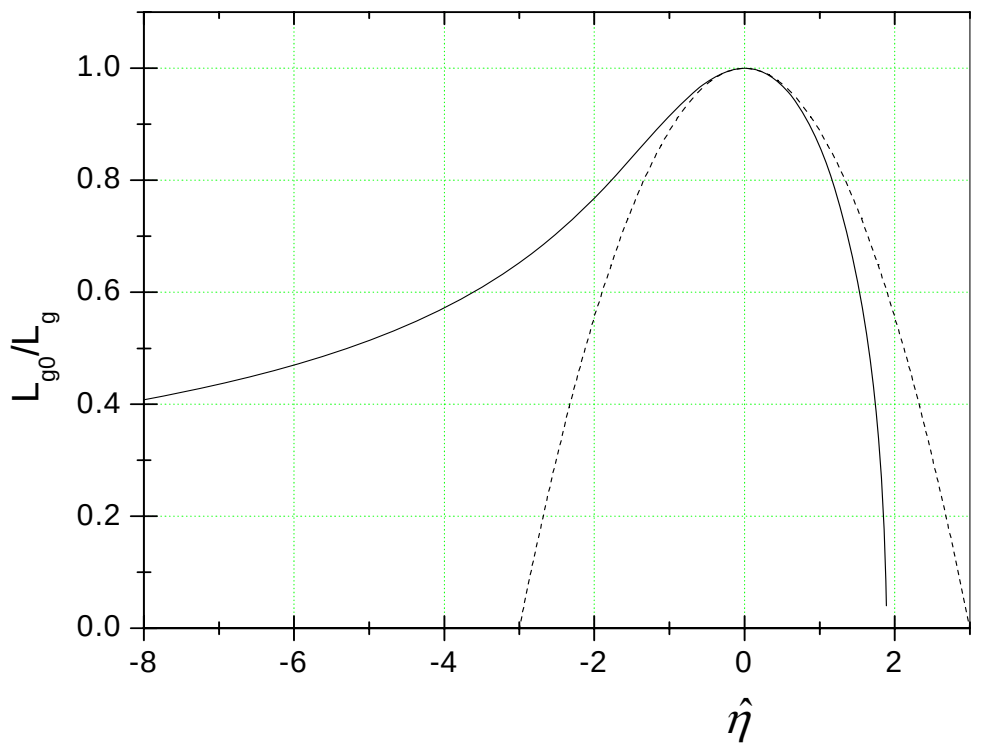

Figure 1, The gain length *vs*. the detuning parameter (Eq.8).
The dashed line is given by Eq.(5)

Furhermore, we give a simplified approximation formula for the exponential gain calculation in the non-resonant case as following:

$$\frac{P}{P_0} \approx \frac{1}{9}(1+\frac{\hat{\eta}}{3})^2 e^{z/L_g}, \ (-3<\hat{\eta}<2, \ z>5L_g) \tag{9}$$

$$L_g \approx L_{g0}/[1-(\frac{\hat{\eta}}{3})^2-\frac{2}{3}(\frac{\hat{\eta}}{3})^3] \tag{10}$$

It agrees well with the exact result from Eq.(7,8) and Eq.(1) (Figure 2). If the cubic term in Eq.(10) is neglectd, it becomes the result of Eq.(5).

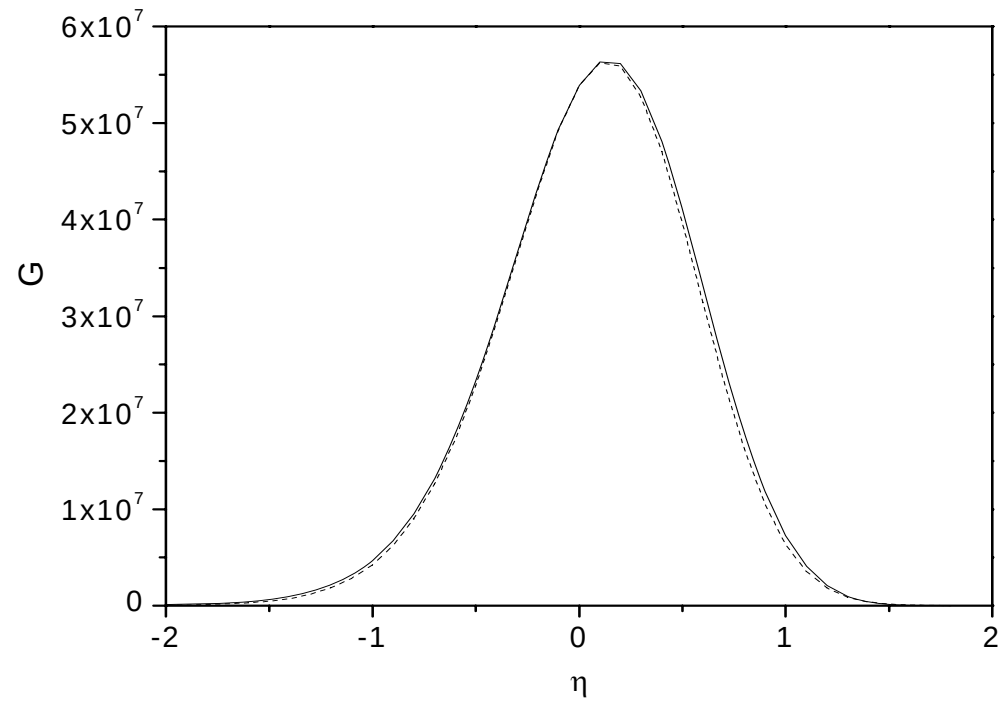

Figure 2, The variation of the gain with the detuning parameter (z= $20L_g$)The solid line: Eq.(7) and (1); the dashed line: Eq.(9)

## IV, INFLUENCE OF BEAM ENERGY SPREAD

For the electron beam having an initial energy spread, the situation is more complex, the characteristic equation (the dispersion relation) now has the form as

$$\hat{\mu} - i\int \frac{f_0}{(\hat{\mu}+i\hat{\eta})^2} d\hat{\eta} = 0 \tag{11}$$

where $f_0$ is normalized distribution of the initial detuning parameter. Generally, the dispersion relation has to be solved numerically for a given initial detuning distribution, the analytic solutions exist only for few types of distributions. Here we consider a simple case, the rectangular distribution with a half-width $\Delta\hat{\eta} = \delta$

$$f(\hat{\eta}) = 1/2\delta, \quad \hat{\eta}_m - \delta < \hat{\eta} < \hat{\eta}_m + \delta$$

Then the cubic equation becomes as

$$\hat{\mu}[(\hat{\mu}+i\hat{\eta}_m)^2+\delta^2]=i \tag{12}$$

The corresponding optical field expression Eq.(1) becomes as

$$\tilde{a}_s(z)=\tilde{a}_{s0}\sum_{j=1}^{3}e^{\hat{\mu}_j\hat{z}}\frac{(\hat{\mu}_j+i\hat{\eta}_m)^2+\delta^2}{(3\hat{\mu}_j+i\hat{\eta}_m)(\hat{\mu}_j+i\hat{\eta}_m)+\delta^2}=\tilde{a}_{s0}\sum_{j=1}^{3}\frac{ie^{\hat{\mu}_j\hat{z}}}{\hat{\mu}_j[(3\hat{\mu}_j+i\hat{\eta}_m)(\hat{\mu}_j+i\hat{\eta}_m)+\delta^2]} \tag{13}$$

Solving Eq.( 12) , we give its solution which has the same form as Eq.(7) but the $p,q$ in it are replaced with

$$p,q=\frac{1}{\sqrt[3]{2}L_{g0}}\{1-2(\frac{\hat{\eta}_m}{3})^3+\frac{2}{3}\hat{\eta}_m\delta^2\pm\sqrt{[1-2(\frac{\hat{\eta}_m}{3})^3+\frac{2}{3}\hat{\eta}_m\delta^2]^2-4[(\frac{\hat{\eta}_m}{3})^2+\delta^2]^3}\}^{\frac{1}{3}} \tag{14}$$

From the root corresponding the growing mode, the gain length as the function of the detuning can be given. In Figure 3 the gain length versus $\hat{\eta}_m$ is plotted by the formulas for different values of the energy spread δ. The curves are the same as that numerically given in Ref.[1]. It shows that as the energy spread increases, the gain length increases, and the detuning parameter corresponding to the shortest gain length also increases.

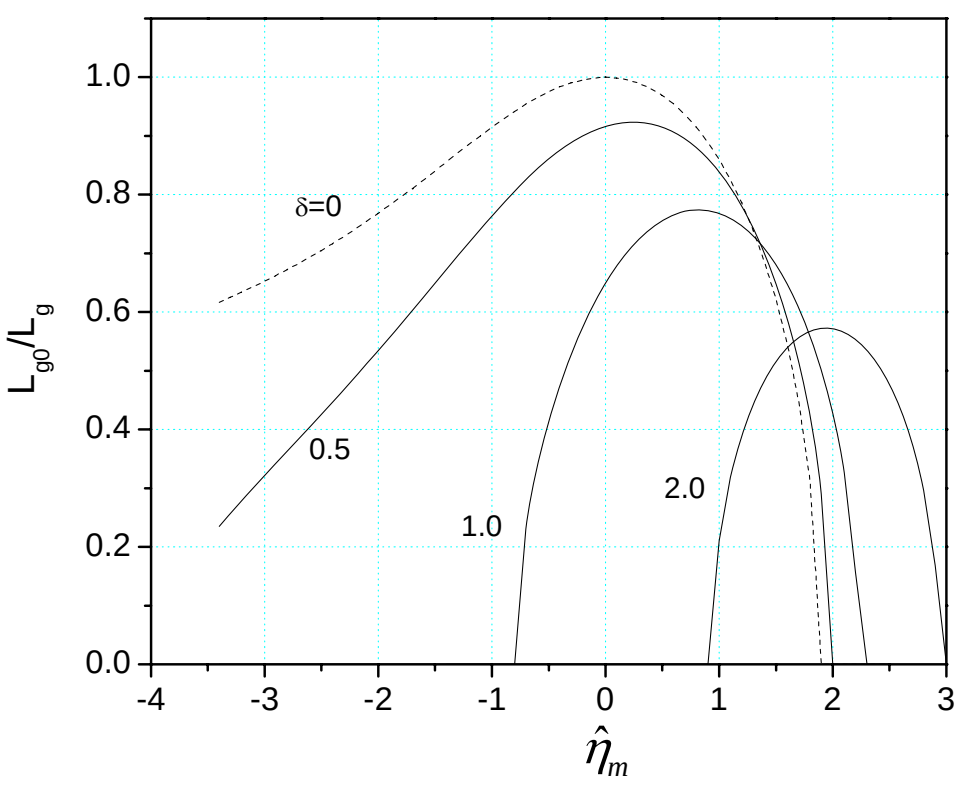

Figure 3, Energy spread effect on the gain length for a rectangular energy distribution with different half-width δ.

The optical field gain for a rectangular distribution of the beam energy can be calculated by Eq.(13). Figure 4 shows the influence of the energy spread on the gain bandwidth near the saturation. As the energy spread increase, the detuning parameter corresponding to the maximum gain also increases, and the gain bandwidth becomes a little bit narrower. The optical power evolutions for different energy spread are calculated and shown in Figure5. We can see that comparing with the mono-energetic beam case achieved optical power dropped about two orders for an energy spread of δ=0.5

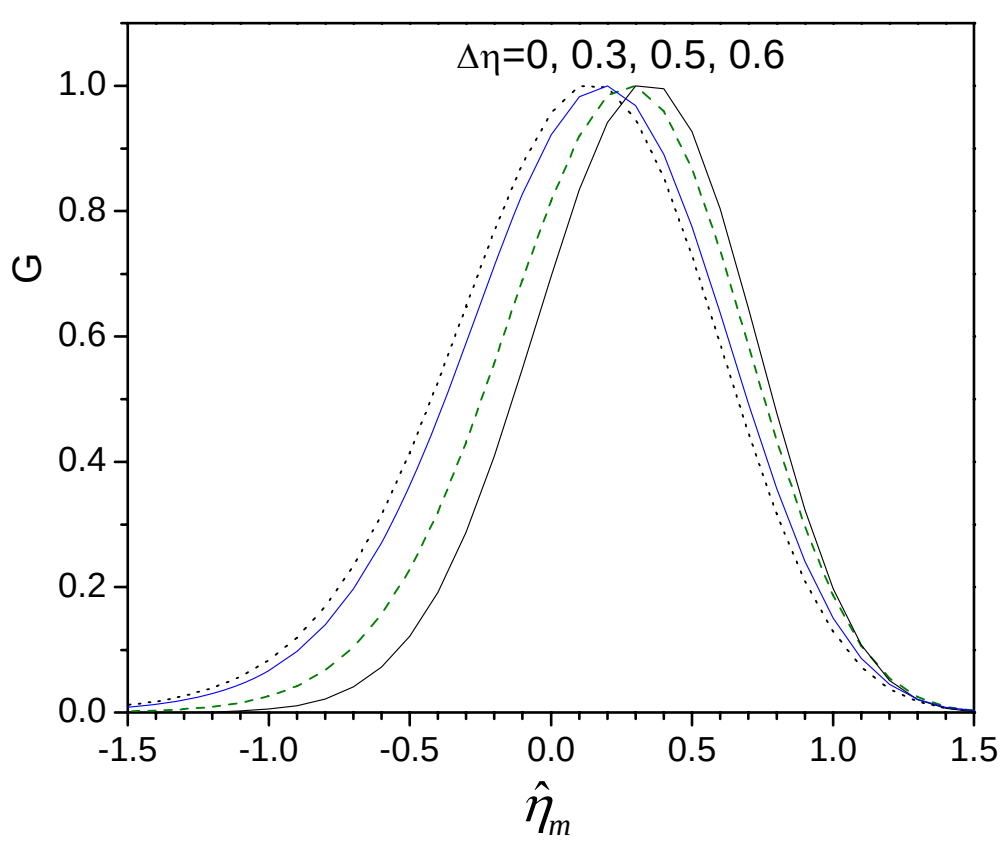

Figure 4，Normalized gain vs. the detuning parameter for different energy spread, (rectangular energy distribution)

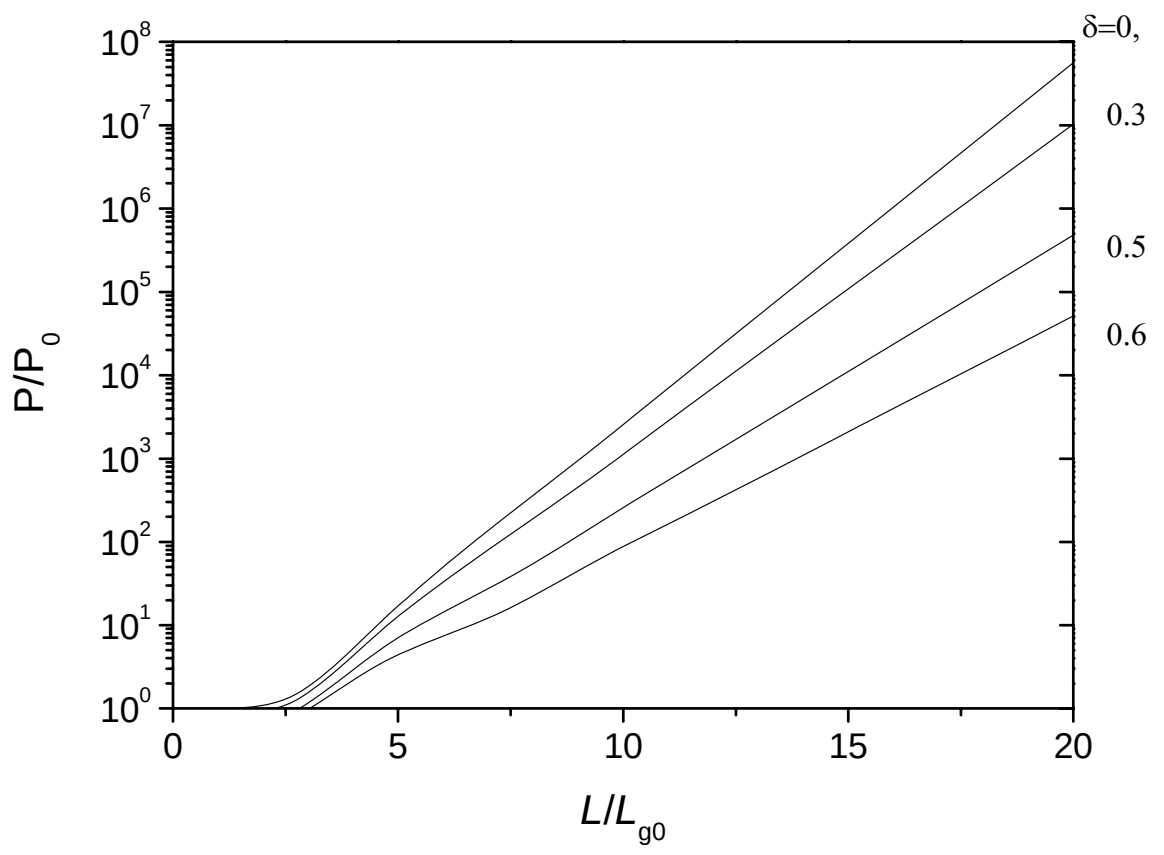

Figure 5，The optical power evolution with different energy spread (rectangular energy distribution)

In addition, we find that the characteristic cubic equation including the impact of the space charge for a mono-energetic beam

$$\hat{\mu}[(\hat{\mu}+i\hat{\eta})^2+\widehat{k}_p^2]=i \tag{15}$$

is the same as that for a beam with rectangular energy distribution (Eq.12), only the $\delta^2$ in it is substituted by $\widehat{k}_p^2=k_p^2/(2k_u\rho)^2, k_p^2=\lambda_s\omega_p^2/\lambda_u c^2\gamma$. Therefore the impact of the space charge on the gain also can be explicitly given by corresponding substitution in Eq.13-14 and Fig.3-5. We obtain the requirement of the practic physical quantities for neglecting the space charge field from $\widehat{k}_p^2<1$

$$\rho<\frac{a_u^2 JJ^2}{2(1+a_u^2)} \tag{16}$$

where $a_u$ is dimensionless vector potential of the *rms* undulator field, *JJ* is the usual Bessel function factor.

Besides rectangular energy distribution, the analytic solution also exist for the Lorentz distribution

$$f(\hat{\eta})=\frac{1}{\pi}\frac{\delta}{(\hat{\eta}-\hat{\eta}_m)^2+\delta^2} \tag{17}$$

here $\delta=\sigma_\gamma/\gamma_0\rho$ is normalized energy spread. The cubic characteristic equation now is

$$\hat{\mu}(\hat{\mu}+i\hat{\eta}_m+\delta)^2=i \tag{18}$$

compare it to Eq.(2), we can give its explicit solution in a similar way, only need to replace the $\hat{\eta}$ in Eq.(2) with $\hat{\eta}_m-i\delta$.

## V, FROM THE LOW GAIN TO THE HIGH GAIN

From the expressions of the optical field and the characteristic roots (Eq.1 and Eq.7), not only the exponential gain, the gain for general case from the low gain to the high gain can be given. The transition from the low gain to the high gain was involved in several publications (e.g. in Ref.[1],[2]), here we give an analysis on it in detail. The level of FEL gain is determined by the number of the gain length included in the undulator. In the low gain regime, the small signal gain formula is often used:

$$g_{ss}=-\hat{z}^3\frac{\partial}{\partial x}\sin c^2\frac{x}{2},\ (x=\hat{\eta}\hat{z}) \tag{19}$$

The comparison between it and the exact result is presented in Figure 6. It shows when the length of undulator is larger than two gain length, the deviation of the small signal gain formula becomes large.

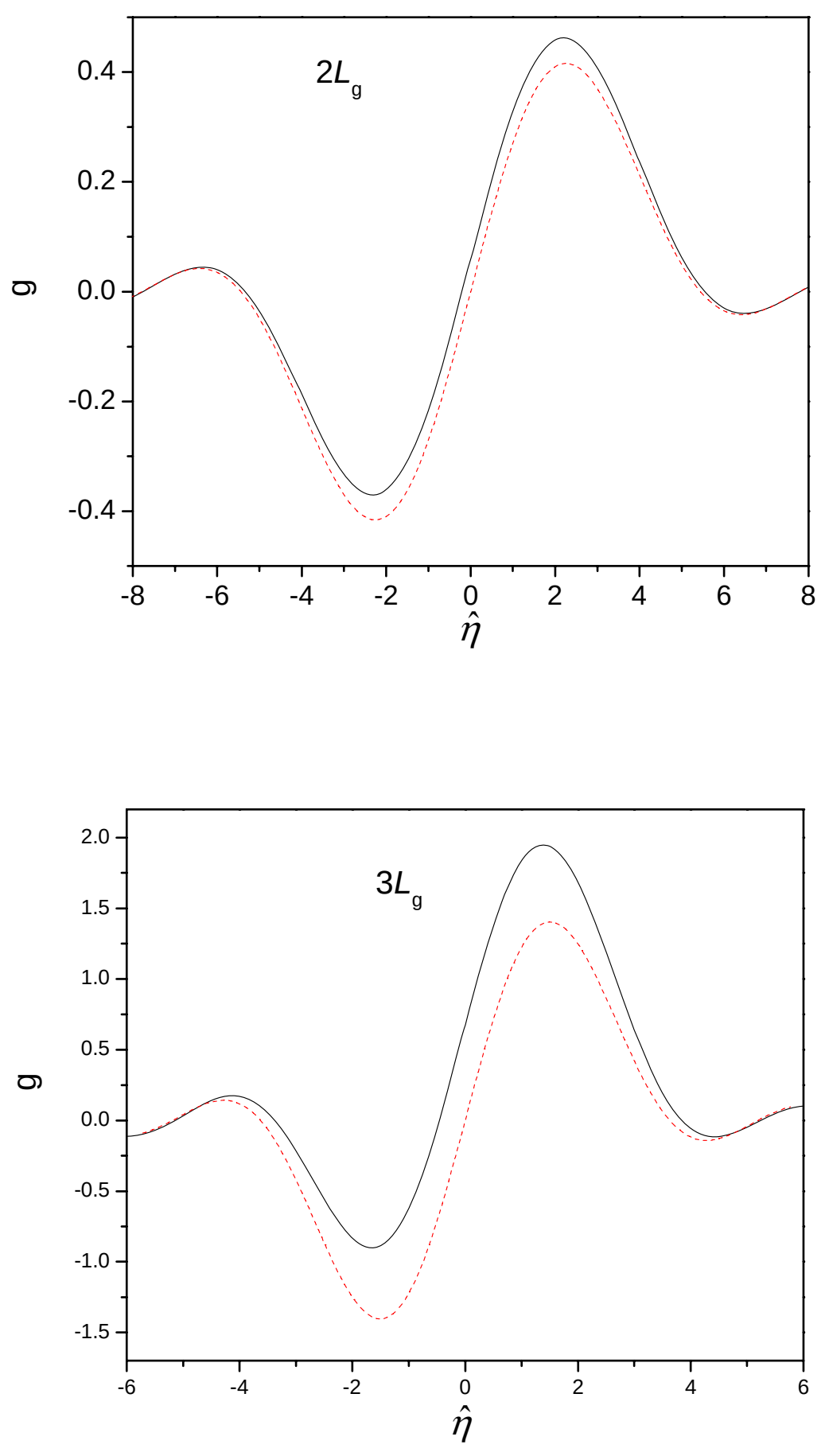

Figure 6, Low gain, the exact result (from Eq.7 and Eq.1) and the result of the small signal gain formula (Eq.19, dashed line)

The variation of the gain with the detuning parameter for several different length of undulator is shown in Figure 7. From Figure 6 and Figure 7, it can be seen that as the length of undulator increases, the gain bandwidth decreases,
from $\Delta\gamma/\gamma \approx 1/2N$ for the small signal gain to $\Delta\gamma/\gamma \approx \rho$ for the exponential gain, and the detuning parameter corresponding to the maximum gain decreases from $\hat{\eta} \approx 4.5L_g/L$ for the small signal gain to $\hat{\eta} \approx 0.12$ (~$2.6L_g/L$) for the exponential gain

The variation of the maximum gain with the undulator length is calculated and plotted in Figure 8, For comparison, the gain at the resonance (Eq.3), the exponential gain (the approximation of Eq.3) and the small signal gain (Eq.19) are also plotted in Figure 8. We can find that the small signal gain formula is more accurate for the case of $L < 2L_g$, while the exponential gain formula is more accurate for the case of $L > 5\ L_g$, the gain curves of two formulas cross at $L/L_g$ ~3.2..

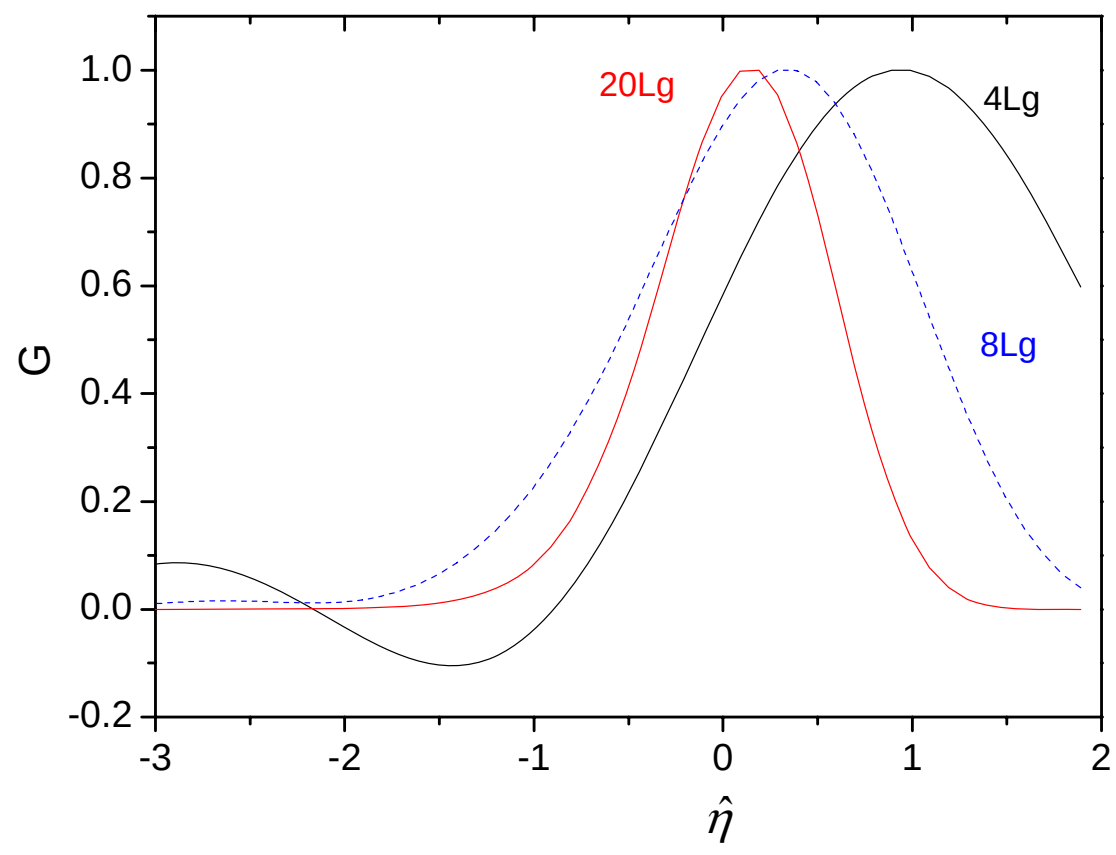

Figure 7， The variation of the normanized gain with the detuning parameter for different number of gain length.

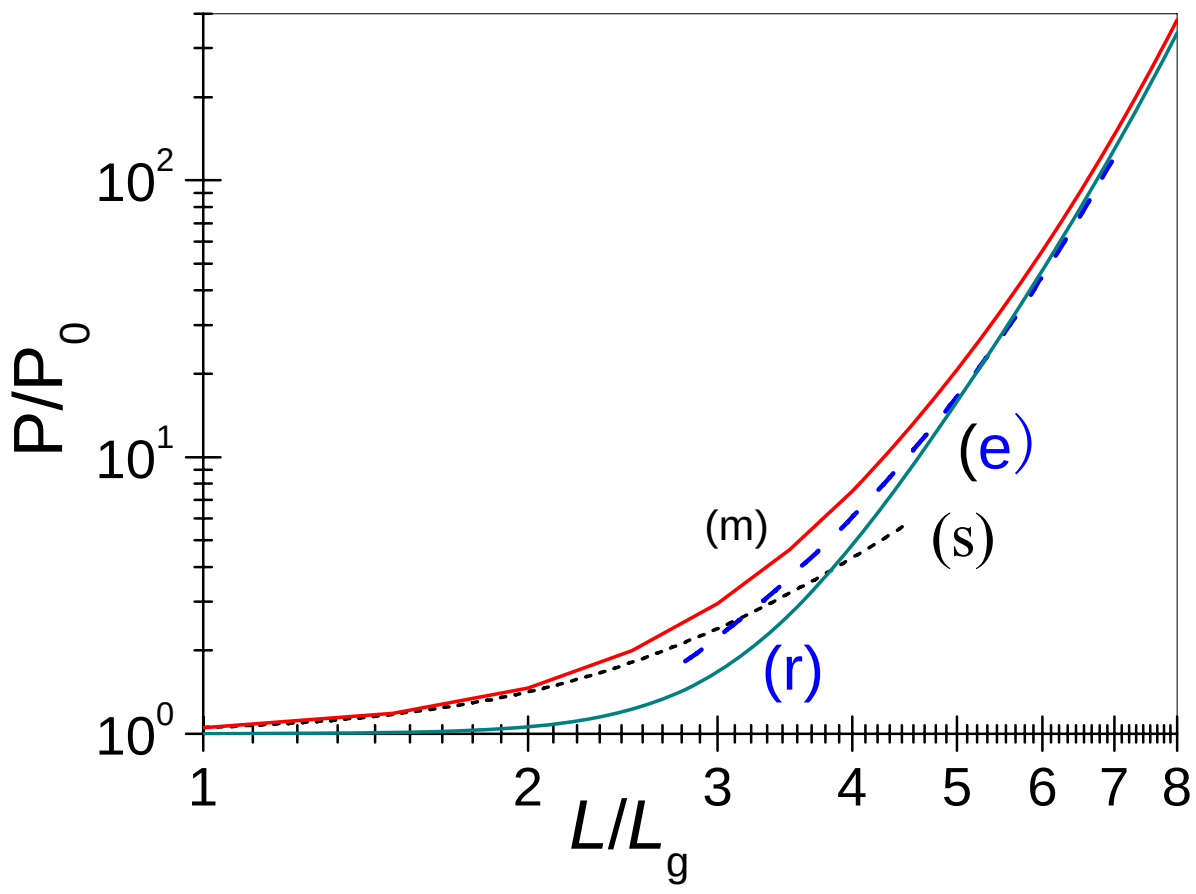

Figure 8, The gain vs undulator length.

The solid line (m): the maximum gain; The dotted line (s): the small signal gain formula(Eq.19); The solid line (r): the gain at the resonance (Eq.3); The dashed line (e): the exponential gain formula (the approximation of Eq.3). Notice that for the lines of the maximum gain and the small signal gain, the detunings are dependent on the longitudinal position.

## VI, SUMMARY

We have studied the FEL gain formulas for non-resonant case. For the non-resonant electron beam, we give the explicit solution of the FEL characteristic cubic equation with a form much more simple than in the literatures ([1],[6])， and explicitly give the exact expression of the gain length as the function of the detuning parameter. We also give a simplified approximation formula for the exponential gain calculation in the non-resonant case. Then one can calculate the gain easily for different detuning parameter and from low to high, and calculate the gain length of general case conveniently.

For the case of the electron beam having an initial energy spread, we give the solution of the characteristic cubic equation explicitly for rectangular energy distribution and Lorentz distribution, respectively. Moreover we find that the explicit solution also can be used for the characteristic cubic equation including the impact of space charge, and give the requirement for the space charge field to be neglected.

We analyzed the transition of the gain from low to high. As the number of the gain length increased, the gain changes from the small signal gain to the exponential gain, the gain bandwidth decreases, from $\Delta\gamma/\gamma\approx 1/2N$ to $\Delta\gamma/\gamma\approx\rho$, and the detuning parameter corresponding to the maximum gain decreases from $\delta\gamma/\gamma\approx$ $26/2k_uL$ to $\delta\gamma/\gamma \approx 0.12\rho$, though the maximum exponential growth rate is at the resonance $\delta\gamma/\gamma=0$. The relations of the different gain formulas are revealed. It is shown that the small signal gain formula is more accurate in the case of $L < 2L_g$, while the exponential gain formula is more accurate in the case of $L > 5\ L_g$. Roughly, the small signal gain formula can be used for the undulator length smaller than about three gain length, and conversely, the exponential gain formula can be used.

The obtained analytical results provide convenience for the gain calculation, will help to the analysis and design of a FEL experiment, and also help to develop insights into the FEL gain process.

**ACKNOWLEDGMENTS**

This work is partly supported by the Major State Basic Research Development Programme of China under Grant No. 2011CB808301 and the National Nature Science Foundation of China under Grant No. 11375199

**REFERENCES**

[1] R. BONIFACIO, et al., Physics of the High-Gain FEL and Superradiance. RIVISTA DEL NUOVO CIMENTO VOL. 13, N. 9 1990

[2] P. Schmuser, et al., Ultraviolet and Soft X-Ray Free-Electron Lasers: Introduction to Physical Principles, Experimental Results, Technological Challenges, STMP 229 (Springer, Berlin Heidelberg 2008), DOI 10.1007/ 978-3-540-79572-8

[3] Zhirong Huang and Kwang-Je Kim, Review of x-ray free-electron laser theory, PHYSICAL REVIEW SPECIAL TOPICS - ACCELERATORS AND BEAMS 10, 034801 (2007)

[4] R. Bonifacio, C. Pellegrini, and L. M. Narducci, Opt.Commun.50, 373 (1984).

[5] Kwang-Je Kim, Nuclear Instruments and Methods in Physics Research Section A, Volume 250, Issues 1–2, 1 September 1986, Pages 396–403.

[6] G. Dattoli, M. Del Franco, M. Labat, P. L. Ottaviani and S. Pagnutti (2012). Introduction to the Physics of Free Electron Laser and Comparison with Conventional Laser Sources, Free Electron Lasers, Dr. Sandor Varro(Ed.), ISBN: 978-953-51-0279-3, InTech, Available from: http://www.intechopen.com/books/free-electron-lasers/free-electron-laser-devices-a-comparison-with-ordinary-laser-source

[7] E.L. Saldin, E.A. Schneidmiller and M.V. Yurkov, The physics of free electron lasers. An introduction, Physics Reports 260 ( 1995) 187-327